\documentclass[12pt]{iopart}

\usepackage{epsfig}
\usepackage{amssymb}
\usepackage{latexsym}

\newcommand{\beq}{\begin{equation}}
\newcommand{\eeq}{\end{equation}}
\newcommand{\bea}{\begin{eqnarray}}
\newcommand{\eea}{\end{eqnarray}}

\begin{document}

\title[]{Intrinsic limitations of inverse inference in the pairwise
  Ising spin glass}

\author{Enzo Marinari and Valery Van Kerrebroeck}

\address{Dipartimento di Fisica, INFM-CNR ed INFN, 
Universit\'a di Roma ``La Sapienza'', 
P. A. Moro 2, 00185 Roma, Italy}
\ead{valery.vankerrebroeck@roma1.infn.it}
\begin{abstract}
  We analyze the limits inherent to the inverse reconstruction of a
  pairwise Ising spin glass based on susceptibility propagation.
  We establish the conditions under which the susceptibility
  propagation algorithm is able to reconstruct the characteristics of
  the network given first- and second-order local observables,
  evaluate eventual errors due to various types of noise in the
  originally observed data, and discuss the scaling of the problem
  with the number of degrees of freedom.
\end{abstract}

\maketitle

\section{Introduction}

The problem of inverse inference has been for a long time one of the
main issues in neural networks
analysis~\cite{Bialek2006,Bialek2007,Bolle2009}.  Given a number of
stimuli, one measures the activity of some local components, such as
spike trains of the neurons, to identify which connections between
them, i.e. which synapses, are active, and, possibly, their
strength~\cite{Amit,KappenRodriguez1997}.  More recently the problem
of inverse inference has manifested itself in other branches of
biology. For example in structural biology the problem comes down to
determine the probability distribution of amino acid strings by
observing the way in which proteins naturally
fold~\cite{Socolich.etal-2005}, or, in systems biology it consists in
recovering structural details of protein-protein interactions from
primary sequence information of gene regulatory
networks~\cite{Szallasi-1999,Weigt-PNAS2009}.  A solution to these
types of problems usually comes about in the following way: one
proposes a model capturing the characteristics of the network under
study, and eventually develops methods to retrieve the structural
characteristics, which can allow to disprove the underlying
hypothesis depending on whether the findings are consistent with the observed
behavior.

The pairwise Ising spin glass has been widely used as a starting point
for analyzing the above problems.  The fact that these problems have a
number of common features has allowed to develop several algorithms
addressing the corresponding inverse inference
problem~\cite{SessakMonasson2008,MezardMora2008,Roudi-Hertz2009,Roudi-Aurell2009}.
However, as more accurate methods come along, new issues regarding the
validation of the underlying hypothesis have been raised.  For
instance, it was pointed out in \cite{Roudi-Nirenberg2009} that by
only observing a subset of the nodes composing a neural network,
reconstructing the original couplings according to a pairwise model
does not necessarily lead to any added information as there exists a
one-to-one relation between the couplings and the (normalized) second
order correlation coefficient for a certain fraction of hidden
variables onward. Thus, in this case one needs to consider higher
order couplings.  Moreover, binning spikes of the neurons, and
accordingly representing their status by Ising-like variables, can
oversimplify the actual observations. Similarly, determining the
possible states in protein folding needs at least Potts-like
variables.  Note also that while for protein networks it is natural to
assume a pairwise interaction model, in neural networks this is not
necessarily the case and though in theory any network can always be
transformed into one which contains only pairwise
interactions~\cite{YedidiaFreemanWeiss-2002}, the practical way to go
about this might not always be obvious.

In this paper, we do not address the interesting cases of the presence
of hidden variables, of higher order interactions or of Potts-like
variables.  Rather, we question a more basic issue: given a situation
where the pairwise Ising spin glass correctly describes the structure of
the network under study, how much information can be gathered from an
experimental input about the values of the couplings of the model?  In
other words, given all two-point correlations, or equivalently, given
the susceptibilities, up to which point can we say something about the
reconstructed couplings?  Does this information allow us to
reconstruct the original model or are there intrinsic uncertainties to
this inverse inference problem? How does the quality of the
reconstruction depend on the original distribution of couplings or on
the size of the system?  The main question can be phrased as to
what amount do the statistical errors that affect the measurements
inhibit us to reconstruct the original model: if the original
observations are incomplete or noisy, up to which point does it make
sense to try and reconstruct the data?  Ideally, one would like to
answer the above questions from a theoretical viewpoint. Here,
however, we start by numerically investigating several of the above
problems by means of a message  passing algorithm, first introduced
in~\cite{MezardMora2008}, which is currently among those delivering
the best results~\cite{SessakMonasson2008}.  We will use this
algorithm to analyze the reconstruction of various types of networks
given their first- and second-order local, possibly noisy,
observables.

We want to analyze some basic features of the
reconstruction process, by spotting some relevant weakness and by
trying to focus on systematic trends that can be relevant in
exploiting this approach and, consequently, in trying to devise improvements 
that could lead to better performances. We consider these result as a
toolbox spelling and clarify a number of facts that can be useful
for a better understanding of and improving this class of methods.

In Section~\ref{S:MET} we introduce the message passing method, and
define some relevant quantities.  We describe in detail the iterative
rules in presence of a memory term, that allows the convergence to a
fixed point.  In Section~\ref{S:DIS} we analyze the reconstruction
procedure for different distribution of the couplings: we look at
binary random couplings and at Gaussian couplings.  In
Section~\ref{S:SYN} we introduce synthetic random errors, by modifying
randomly exact values for the susceptibility, and we try to understand
how a larger incertitude affects the quality of the reconstruction of
the couplings. We analyze both the case of an additive error and the
one of a multiplicative error.  In Section~\ref{S:MC} we analyze data
obtained by a Monte Carlo simulation, and we study the quality of the
reconstruction as a function of the accuracy of the measurements. We
draw our conclusions in Section~\ref{S:CON}.

\section{Susceptibility propagation and the inverse Ising spin glass\label{S:MET}}

While message passing algorithms have been widely used to solve the
direct inference problem where the characteristics of the underlying
network are given and one wants to derive experimentally observable
quantities~\cite{MezardParisi2000}, their adaptation to tackle the
inverse problem is relatively recent.  Here we consider the inverse
Ising spin glass, which assumes that the basic constituent agents of the
network interact only in a pairwise, symmetric way with the other
agents.  In other words, we assume the problem is described by the
following partition function:
\begin{equation}
Z=\sum_{\mathbf{\sigma}}\exp \left[-\frac{1}{T} 
\left(\sum_{i=1}^N h_i \sigma_i + \sum_{i<j} J_{ij} \sigma_i \sigma_j \right)
\right]\;,
\label{eq:part_fction}
\end{equation}
where the $\sigma_i$ can take the values $\pm 1$, $h_i$ are
(potentially site dependent) magnetic fields, the couplings $J_{ij}$
are quenched random variables that can have both a positive and a
negative sign and are distributed under a probability function that we
will discuss in the following, and $T$ is a ``temperature'' governing
the behavior of the system.  We can now define the inverse Ising spin
glass problem by considering as given (by a, potentially numerical,
experiment) the local magnetizations and the susceptibilities, and by
trying to compute the $N$ local fields $h_i$ and the $N(N-1)/2$
couplings $J_{ij}$.

M\'ezard and Mora have recently introduced a message passing
procedure, the so called {\it susceptibility
  propagation}~\cite{MezardMora2008}, that is able, in appropriate
conditions, to solve this problem.  As for the direct problem, it is a
distributed algorithm where messages are exchanged between any two
pairs of nodes. In case of the inverse problem, the content of some of
these messages concerns the probability distribution of the local
field of a vertex, while other messages contain information about the
distribution of the couplings between two vertexes.

More specifically in the inverse problem one defines four types of
messages, $h_{i \to j}$, $u_{i \to j}$, $v_{i \to j,k}$ and $g_{i \to
  j,k}$.  The messages $h_{i \to j}$ and $u_{i \to j}$ are exchanged
between couples of nodes, while the messages $v_{i \to j,k}$ and $g_{i
  \to j,k}$ are exchanged between triples of nodes.  The
susceptibility propagation algorithm starts by assigning a random
value to each of these messages and setting all the ``estimated
couplings'' $J_{ij}$ to zero. These quantities are updated iteratively
according to the following rules (where the $m_i$ and the $\chi_{ij}$
are the ``experimental'' inputs):
\begin{eqnarray}
h_{i \to j} & = & \rm{arctanh}\left( m_i\right) -u_{j \to i} 
\label{eq:update_rule_h}\\
g_{i \to j,k} & = &\frac{\delta_{i,k}}{T} + 
\sum_{l \in \partial i\setminus j} v_{l \to i,k} \label{eq:update_rule_g}\\
\tanh \left(\frac{J_{ij}}{T}\right) & = & 
\varepsilon \left[ \frac{\tilde{C}_{ij} 
- \tanh \left(h_{i \to j}\right) 
\tanh \left(h_{j \to i}\right) }
{1- \tilde{C}_{ij}\tanh \left(h_{i \to j}\right) 
\tanh \left(h_{j \to i} \right)}\right]\\ 
&+& \left(1-\varepsilon\right) 
\tanh \left(\frac{J_{ij}}{T}\right) \label{eq:update_rule_j}\\
\tanh \left(u_{k \to i}\right) & = & 
\tanh \left(\frac{J_{ik}}{T}\right) \tanh \left(h_{k \to i}\right) 
\label{eq:update_rule_u}\\
v_{l \to i,k} & = & g_{l \to i,k} 
\tanh \left(\frac{J_{il}}{T} \right)
\frac{1 - \tanh^2 \left(h_{l \to i}\right)}
{1 - \tanh^2 \left(u_{l \to i}\right)}\;,
\label{eq:update_rule_v}
\end{eqnarray} 
where
\begin{equation}
\nonumber
\tilde{C}_{ij}  \equiv \frac{\chi_{ij}-g_{i \to j,j}(1-m_i^2)}{g_{j\to i,j}} 
+ m_i m_j \label{eq:update_rule_c}\;.
\end{equation}
For a detailed discussion of these equations (with $\varepsilon=0$),
see~\cite{Mora2007}.  If this set of equations converges to a fixed
point, the local fields are reconstructed as
\begin{equation}
h_i  = T \left( \rm{arctanh}\left( m_i\right) 
- \sum_{k \in \partial i} u_{k \to i}\right)\;.
\label{eq:local_fields}
\end{equation}
The couplings can be read directly from equation
(\ref{eq:update_rule_j}).

The additional external parameter~\cite{MezardMoraPC}
$\varepsilon\in]0,1[$ that we have introduced in the update rule
(\ref{eq:update_rule_j}) is a memory term: we keep, when computing
 the new value of a coupling, part of the old value, and only modify
 the coupling partially. This is a possible way out to the fact that
 for $\varepsilon=0$ the iteration typically fails since it ends up to
 propose an absolute value larger than one for updating
 $\tanh\left(J_{ij}/T\right)$. Clearly this slows down the
 convergence: in our scheme we had to use very small values of
 $\varepsilon$, but this eventually guaranteed that the susceptibility
 propagation algorithm was eventually leading in most part of the
 cases to a fixed point solution (as we discuss in the following
 section).  We will discuss in some more detail the role of the
 ``temperature'' $T$ that appears in an explicit form in our iterative
 equations: if one thinks the local susceptibilities to be connected
 to spin-spin correlation functions according to the usual relation
 $\chi_{ij}=\beta_P \langle \sigma_i \sigma_j\rangle$ (where $\beta_P$
 is the ``physical'' value of the inverse temperature) the parameter
 $T$ is naturally related to $\beta_P^{-1}$, but one has indeed some
 more freedom in tuning $T$.

In this note we will consider the case where all local fields are zero
and we will focus on how to reconstruct the couplings when the local
susceptibilities are given. One way to evaluate how close the
reconstructed couplings are to the original ones is based on computing the
Kullback-Leibler distance, which measures the weighted difference
between the exact and the reconstructed
distribution~\cite{KappenRodriguez1997,Roudi-Nirenberg2009}.  An exact
computation of the Kullback-Leibler distance involves a sum over all
the states of the system, and can only be performed for rather small
systems. We have used for our comparison a simpler indicator, that is
appropriate for our goal:
we
consider the ratio of the
average of the squared difference of the
reconstructed couplings
$J_{ij}^r$ and the original couplings $J_{ij}$, and the variance
$\sigma$ of the distribution of the exact couplings,
\begin{equation}
\Delta=\frac{\sqrt{ \left\langle \frac{\sum_{i<j}(J_{ij}^{r}-J_{ij})^2
    }{N(N-1)/2}\right\rangle}} {\sigma}\;.
\label{eq:error}
\end{equation}
In the following we always consider coupling distributions with zero
mean and variance $\sigma=\tilde{J}/\sqrt{N}$, and we will set
$\tilde{J}\equiv 1$. In the case of fully connected graphs, i.e. the
usual mean field theory of spin glasses (Sherrington-Kirkpatrick
model), the spin glass phase sets in at
$T_c = 1$~\cite{SherringtonKirkpatrick1975}.
Using a different terminology~\cite{KappenRodriguez1997,SessakMonasson2008}, 
one fixes the temperature (that disappears from the iteration scheme) 
$T \equiv 1$ and looks at $\tilde{J}^{-1/2}$, in which case the
critical transition on the complete graph occurs at $\tilde{J}=1$.
Considering explicitly
a ``temperature'' 
has the advantage that it provides direct connections to
the (direct) studies of spin glasses in statistical mechanics, and
it can allow  
to use the $T$ parameter to modify the iterative procedure.  We
have always averaged the quantity $\Delta$ defined in
Eq.~(\ref{eq:error}) over either a hundred (for small values of $N$)
or twenty (for larger values $N$) different distributions of couplings
(both in the case of binary couplings and in the case of Gaussian
couplings).  As we will describe in detail in the following we have
analyzed both cases where the input to our inverse reconstruction
procedure was based on exact values of the observables ($m$ and
$\chi$) computed using an exact enumeration in small systems and cases
where they were computed by Monte Carlo simulations (where the
statistical accuracy of the time series was kept under control and is
one of the relevant issues we have investigated).

\section{The quality of the reconstruction as a function of temperature
  and of the coupling distribution\label{S:DIS}}

The first issue that we have analyzed is the quality of the
reconstruction as a function of the temperature $T$ of the Ising spin glass
and of the distribution of the couplings. We consider the two, somehow
extreme cases, of binary couplings, where $J_{ij}=\pm 1$ with
probability $\frac12$, and of Gaussian couplings, where the $J$ have
zero average and unitary variance. We obtain the susceptibilities that
serve as input to the inverse reconstruction by exact enumeration. Thus, in this section we are dealing with inputs that are exact values (but for
the finite precision of the input words, see later). All the results
are sample averages, as described in the former section. 

\begin{figure}[!ht]
\centering
\includegraphics[angle=270, width=0.9\textwidth]{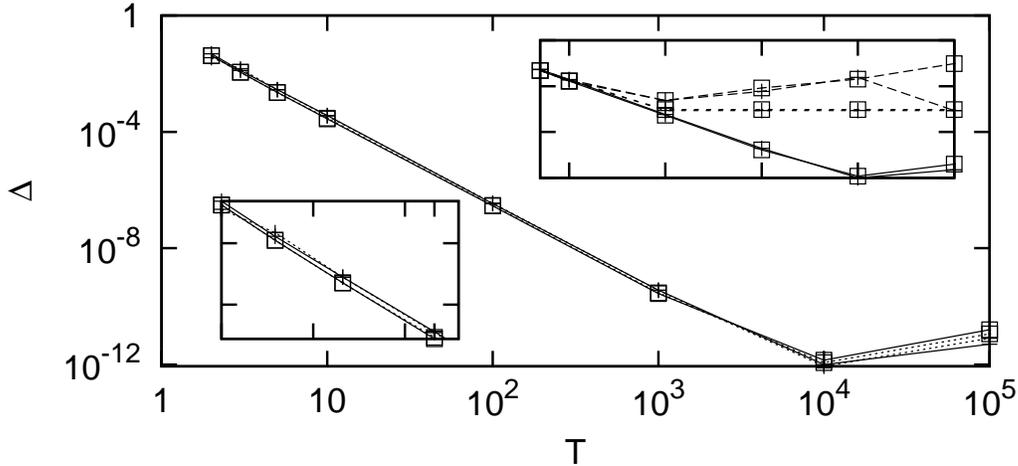}
\caption{\label{fig:exact_distributiondependence} The reconstruction
  error $\Delta$ as a function of $T$ for fully connected graphs with
  binary (full lines) or Gaussian distributed couplings (dotted
  lines): here variables occupy $12$ bytes. 
  We considered two
  different sizes of the graphs: $N=10$ ($+$) and $N=20$
  ($\square$). The $x$ and $y$ scales of the insets can be easily
  deduced from the main plot. The lower inset contains a detail of the
  same plot for $T \in [2,12]$. The upper inset demonstrates the
  dependence of $\Delta(T)$ on the precision of the variables used in
  implementing the susceptibility propagation algorithm that allows
  reconstructing the binary distributed couplings. From top to bottom,
  we used $4$, $8$ and $12$ byte variables, respectively. }
\end{figure}

In Fig.~\ref{fig:exact_distributiondependence} we show a plot of the
average error (\ref{eq:error}) of the reconstruction of the couplings
as a function of the temperature for the two different distributions
of the couplings and for different system sizes (in the main plot of
the figure we used variables of $12$ byte).  With our normalization
$T=1$ is $T_c$, the critical temperature of the statistical model (the
mean field, Sherrington-Kirkpatrick spin glass).

In a large temperature interval the quality of the reconstruction
essentially only depends on the temperature $T$, and not on $N$, nor on
the details of the distribution of the couplings: when increasing $T$
the quality improves.  Close to $T_c$ the error becomes large. Still
even for temperatures of the same order of magnitude as 
$T_c$ (at least down to $T=2$, say) our procedure allows to get precise hints
about the values of the individual couplings.

In the lower inset of the figure we enlarge the left part of the main
plot, for temperatures in the range going from $2$ up to $12$. This 
demonstrates even more clearly that the error in the reconstruction does not
depend much on the type of distribution we considered, nor on the size
of the system. In other words, it is no problem for our algorithm
to distinguish the two different cases where couplings can only take a
few values (here two) and where they come from a continuous
distribution: this can have some relevance when deciding how to
analyze and interpret experimental data.  Also $\Delta$ is an average
of the systematic error of the reconstruction of one single coupling:
this means that the algorithm is essentially more precise for
increasing $N$, as is well known for MP-algorithms and was already
observed in~\cite{MezardMora2008}.

These facts have strong implications about the scope that the pairwise
statistical model we are using can enjoy: there are situations, in
other words, where we cannot gather any useful information by trying
to determine the couplings $J_{ij}$. If all the correlation
functions among the constituent agents of a system are known, and we
know that they only interact in a pairwise manner, we are able to
reconstruct each single coupling if the system is at a relatively high
temperature, or if, in other words, the corresponding distributions of
the couplings have large variance. The fact that for decreasing
temperature the reconstruction becomes less precise, is most likely
due to a lack of information on how to capture the complex structure
of the solution space near and below the critical temperature
$T_c=1$. Unfortunately, all algorithms currently available to tackle
the inverse problem suffer this same symptom.  In the present case of
the susceptibility propagation algorithm the increasingly complex
structure of the phase space also manifests itself as a non-convergent
behavior of the algorithm.  While for larger temperatures, the
algorithm converges to a fixed point, as we approach the critical
point, the average change in the messages
(\ref{eq:update_rule_h})-(\ref{eq:update_rule_v}) between two
consecutive updates gradually converges to some increasing positive
value until they, eventually, no longer converge. 
A better reconstruction 
of the couplings defeating this critical point could use
additional information on the network, or a more advanced 
algorithm which take this extra complexity into account, such as the survey
propagation algorithms~\cite{SP}.

Finally, Fig.~\ref{fig:exact_distributiondependence} evidences another 
important feature, regarding the
numerical precision needed for an accurate reconstruction. If the
correlation functions or the susceptibilities have been obtained at
high temperatures, and we are in the ideal case where at least in
principle we know them exactly, the quality of the reconstruction will
be essentially limited by the precision with which we represent them
and that we use in implementing the iterative scheme.  We analyze this
phenomenon by finding the solution using variables of different sizes
(respectively $4$, $8$ and $12$ bytes) and we plot the results in the
upper inset of Fig.~\ref{fig:exact_distributiondependence}. The
situation is quite clear: selecting a level of precision $p$ sets a
maximum temperature value $T^*(p)$ that would allow to improve the
reconstruction. For $T>T^*(p)$ the information we gather is hidden by
the insufficient precision $p$, and the reconstruction quality does
not improve.  All the results we will discuss in the following have been
obtained with $12$ bytes wide variables.

\section{Synthetic noisy data\label{S:SYN}}

The susceptibilities (or the correlation functions) one obtains as the
output of an experiment are far from exact. The error can either be
due to the limitations of the experimental set-up, thus imposing some
absolute error on the measured date, or to statistical 
fluctuations, that can originate from 
a number of different causes.  In case the
observables are averages of successive experiments, the two-point
correlation functions are limited by some relative errors which can
possibly be improved by performing more experiments. We will discuss
here how these different types of error can affect the reconstructions
of the couplings.

The case of an error that is on average constant in magnitude
(independently from the size of the observable we are considering) and
the one where its ratio to the signal is constant in magnitude are
indeed very different. In the case of an error that is constant on
average small correlations functions will not give any significant
amount of information: if, for example, for a model endowed with a
Euclidean distance $d$ we expect an exponential decay
with $d$, only the first, larger contributions, will be of use in our
reconstruction, while the smaller ones will be completely hidden by
the noise.

We start again from exact values of the susceptibility that we compute
by exact enumeration, summing all the contributions of the $2^N$ spin
configurations.  We simulate the presence of an absolute error by
including an additive noise term to the exact observables: the
susceptibilities $\chi_{ij}^{A}$ used for reconstruction are given
here by the exact susceptibilities $\chi_{ij}$ with the addition of a
noise term $r_{\eta}$, uniformly drawn from the interval
$[-\eta,+\eta]$, with $\eta > 0$: $\chi_{ij}^{A}=\chi_{ij}+r_{\eta}\;,\;\;
\forall\; ij$.

\begin{figure}[!ht]
\centering
\includegraphics[angle=270, width=0.9\textwidth]
{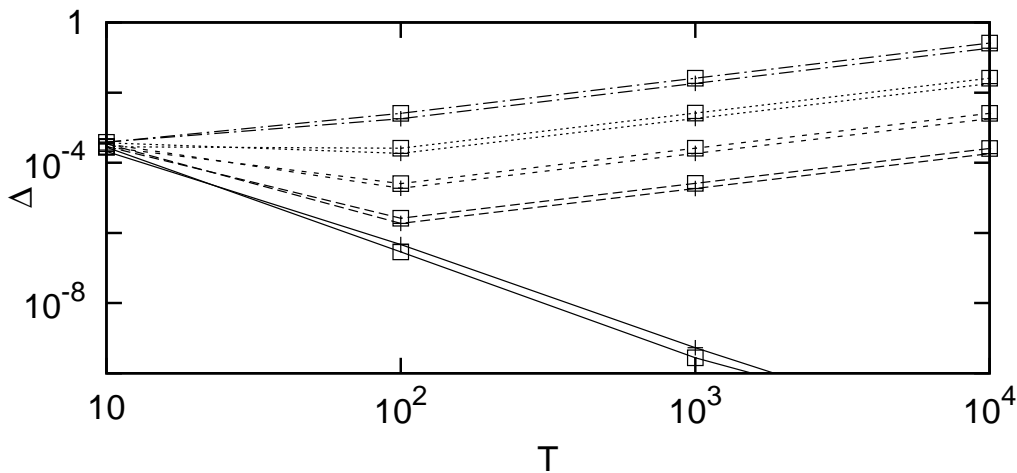}
\caption{\label{fig:ADDnoise} $\Delta$ as a function of $T$. Here the
  susceptibilities used as input for the 
  reconstruction of the couplings are only
  approximate due to an additive noise. From bottom to top:
  $\eta=10^{-8}, 10^{-7}, 10^{-6}, 10^{-5}$ for $N=10$ ($+$) and
  $N=20$ ($\square$).}
\end{figure}

We show our results for the reconstructed couplings in
Fig.~\ref{fig:ADDnoise}.  We show the values obtained for different
choices of $\eta$.  The effect of this random noise is irrelevant for the
low $T$ range where the reconstruction is possible, but becomes large
when $T$ increases. The larger is $\eta$, the smaller is the $T$ range
where the reconstruction becomes unreliable. In presence of this kind
of noise the quality of the reconstruction worsens when $T$ increases:
this can be an interesting observation when trying to optimize a
reconstruction scheme of experimental data.
It is interesting to note that the error on the reconstructed coupling
is several orders of magnitude larger than the additive error on the
susceptibilities.  This is due to the fact that the order of magnitude
of the susceptibilities is large at small temperatures, and much
smaller at high temperatures. Therefore, especially at high
temperatures, additive noise terms are very damaging.

\begin{figure}[!ht]
\centering
\includegraphics[angle=270, width=0.9\textwidth]
{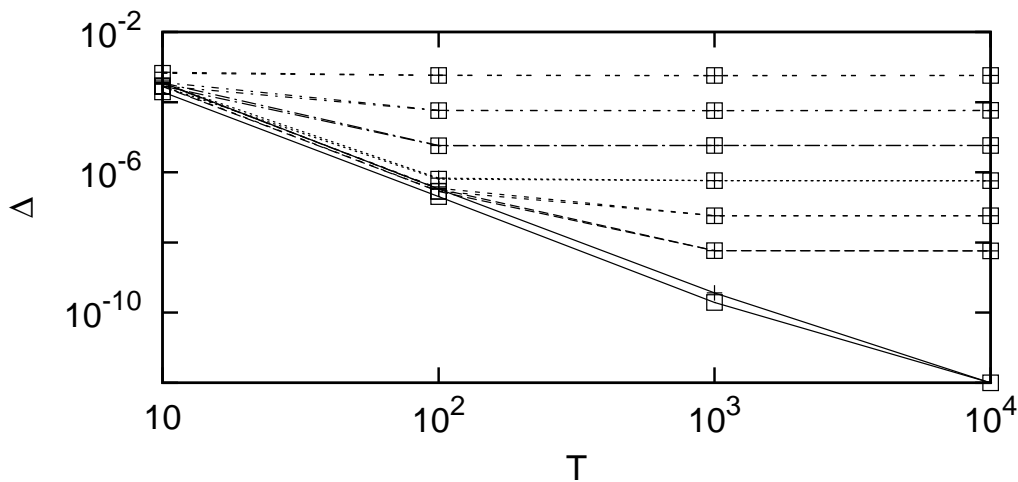}
\caption{\label{fig:MULnoise} $\Delta$ as a function of $T$. Here the
  susceptibilities used as input to the reconstruction scheme are only
  approximate due to the presence of a multiplicative noise. From
  bottom to top: $\epsilon=10^{-8}, 10^{-7}, 10^{-6}, 10^{-5},
  10^{-4}, 10^{-3}$ for $N=10$ ($+$) and $N=20$ ($\square$).}
\end{figure}

In Fig.~\ref{fig:MULnoise} we show the effect of a multiplicative
noise term on the susceptibilities. More precisely, these
reconstructed couplings are computed starting from the approximated
susceptibilities $\chi_{ij}^{M}$, which were obtained from the
original susceptibilities by multiplying them by a factor
$r_{\epsilon}$, which was drawn uniformly from the interval
$[1-\epsilon,1+\epsilon]$, with $\epsilon>0$:
$\chi_{ij}^{'}=r_{\epsilon}\chi_{ij}\;,\;\;\forall\; ij$.  

Again, the error on the observables influences severely the
reconstruction of the couplings at higher temperatures. Here there is
a clearer threshold effect than in the former case, and there is a
clear dependence of the ``breaking point'' $T^*(\eta)$ over
$\eta$. The situation is very similar to the one that we have
discussed in the previous section, where we were using ``short''
variables with a finite, small width (down to four bytes).

\section{Monte Carlo noisy data\label{S:MC}}

A Monte Carlo numerical experiment is one of the best proxy for a real
experiment. One gets sets of data that are asymptotically distributed
according to a certain probability function. These data are affected
by statistical errors, as it would happen in an experiment.  We
analyze here how the reconstruction works when starting from Monte Carlo
data obtained under variable accuracy requirements: this is an issue
of paramount interest, since we need to know if a given real
experiment, with a given level of accuracy, will give information
that can lead to a useful coupling reconstruction.

\begin{figure}[!ht]
\centering
\includegraphics[angle=270, width=0.9\textwidth]
{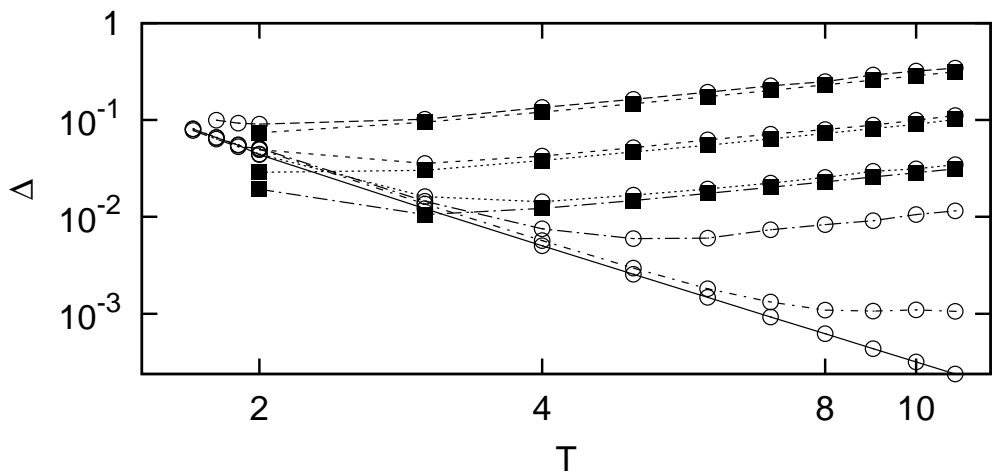}
\caption{\label{fig:MCnoise} The error $\Delta$ as a 
  function of $T$ for 
  fully connected graphs of size $N=16$ ($\circ$)
  and $N=128$ ($\Diamond$), with binary couplings. 
  The different curves represent
  reconstructions starting from the exact susceptibilities for the
  $N=16$ system (continuous line),
  and starting from approximations of the susceptibilities
  generated from $10^4$, $10^5$, $10^6$, $10^7$ and $10^9$ MC data
  (from top to bottom) for $N=16$, and $10^5$, $10^6$ and $10^7$ MC
  data (from top to bottom) for $N=128$.}
\end{figure}

So here we do not start from exact data, but from data obtained by a 
usual Monte Carlo simulation, with a local, accept-reject Metropolis
updating scheme, and we use our inverse algorithm to get the couplings
from these data.  We first lead the system to equilibrium (and discard
data obtained during this thermalization phase of the simulation), and
eventually collect data for a number of Monte Carlo steps.  We show in
Fig.~\ref{fig:MCnoise} the error $\Delta$ on the reconstruction of the
couplings, given the values $\chi_{ij}^{MC}$ of the susceptibilities,
obtained by sampling the solution space with a Monte Carlo Markov
Chain.

Given a fixed number of Monte Carlo measurements the error on the
susceptibilities increases with the temperature, resulting in a less
precise reconstruction of the couplings.  However, by increasing the
duration of the experiments, i.e. increasing the number of independent
observations, the relative error can be drastically reduced as can be
seen from Fig.~\ref{fig:MCnoise}. The pattern is, as one would have
expected, very similar to the one of a statistical error of constant
average size. Indeed, this is exactly what happens here, where we
estimate all correlation functions by adding numbers of order one
(the individual values of the correlations, that can be $\pm 1$).
For each level of the error the reconstruction works as if correlations
were exact up to a given $T$ value, beyond which its accuracy does not
increase anymore with $T$, but, on the contrary, it starts decreasing
with $T$. 

We also analyzed the low-temperature limit down to which the couplings
can be reconstructed starting from approximated two-point
correlations. While the exact correlations in general allow to
reconstruct the couplings down to a temperature as low as $T=1.7$, no
solution could be found, for example, starting from susceptibilities
obtained from only $10^4$ independent MC data at this same
temperature: the susceptibility propagation algorithm can be
additionally limited by an inaccurate original data set.

We have also tried to understand how the performance of the 
susceptibility propagation algorithm varies when we increase the
number of elements of the system. In the Monte Carlo case we have
studied the two cases $N=16$ and $N=128$, where the second system is
eight times larger than the first one: we show both sets of data in
Fig.~\ref{fig:MCnoise}.  Larger size systems require more experiments
to get a reconstruction of the same quality than for the smaller systems:
the $N=128$ curves overlap with 
$N=16$ curves obtained with ten times less statistics. 
After assuming this rescaling
our data clearly show that the reconstruction procedure also works
very well even when we heavily increase the volume of the system.  Let
us look carefully at ``low'' values of $T$.  For example, when starting 
from observables with a good precision, at $T=2$ the
$N=128$ reconstruction clearly improves in quality with respect to those 
for $N=16$ (the
same phenomenon can already be observed, on a smaller scale, in
Fig.~\ref{fig:exact_distributiondependence} when comparing $N=10$ and
$N=20$). Reconstruction on large systems sizes is possible and
reliable even if the ``temperature'' of the system is not so far from
criticality, which certainly is good and useful news.

\section{Conclusions\label{S:CON}}

We have analyzed a number of features of the inverse Ising spin glass
problem, by using the susceptibility propagation algorithm, first
introduced in~\cite{MezardMora2008}. In a very large temperature
window, this algorithm is able to reconstruct the individual couplings
and, consequently, their overall distributions with a remarkable
precision.  If the system is ``at high temperature'' (or, in other,
maybe more physical terms), if the (zero average) disorder does not
fluctuate too much, the quality of the reconstruction is basically
only limited by the precision under which the experimental input is
known, and by the precision used when implementing the susceptibility
propagation algorithm.

For smaller temperatures approaching the critical temperature, or
equivalently, for distributions of the couplings characterized by a
large variance, the reconstruction is less accurate and eventually the
algorithm fails to find any solutions to the problem. This is 
due to the fact that it does not take the possibility of multiple states into account, which is exactly what happens in the spin-glass phase.

All algorithms currently available suffer this same problem. However,
the message passing algorithm used in this paper could possibly be
improved by using the probability distributions of the observables as
basic working ingredients, rather than the observables themselves to
obtain a type of survey propagation algorithm for which the exchanged
messages do not contain information on the couplings, but rather on
the probability distribution of each individual coupling.
Furthermore, the nature of the susceptibility propagation algorithm
suggests it could be easily adapted to include the case of Potts-like
variables allowing to treat problems in structural
biology~\cite{Weigt-PNAS2009}.

While the overall reconstruction of the pairwise model is quite
precise in case the original data set is accurate, the results can
deteriorate fast if data are affected by a statistical error.  The
number of experiments that have to be used to obtain the average
two-point correlations needs to be increasingly large for increasing
sample size. Also, at large temperatures, where the values of the
susceptibilities are small, this error on the reconstruction of the
couplings becomes more pronounced. For the same reason, an absolute
error on the two-point correlations is increasingly damaging at higher
temperatures.

All together we feel that our conclusion lead to an optimistic
scenario. Even in presence of a large statistical or systematic 
ignorance the reconstruction is possible and can be effective. Large
samples still allow for a good reconstruction quality, under the
condition that the statistical inaccuracy that affects the data
is lowered down to a reasonable level.

\section*{Acknowledgments\label{S:ACK}}
We thank Thierry Mora for describing us the use of the
$\varepsilon$ term in this context. We acknowledge interesting
conversations with Federico Ricci-Tersenghi. 



\end{document}